\documentclass[proceedings,article,submit,moreauthors,pdftex]{mdpi} 
\preto{\abstractkeywords}{\nolinenumbers}
\usepackage[flushleft]{threeparttable}

\firstpage{1} 
\pubvolume{xx}
\issuenum{1}
\articlenumber{5}
\pubyear{2019}
\copyrightyear{2019}
\history{Received: date; Accepted: date; Published: date}

\Title{Study of periodic signals from blazars$^\dagger$}

\Author{Gopal Bhatta $^{1,\ddagger}$\orcidA{https://orcid.org/0000-0002-0705-6619}}

\AuthorNames{Firstname Lastname, Firstname Lastname and Firstname Lastname}

\address{%
$^{1}$ \quad Astronomical Observatory, Jagiellonian University, ul. Orla 171, 30-244 Krak\'ow, Poland\\
}

\corres{Correspondence: gopal@oa.uj.edu.pl}

\firstnote{Presented at the meeting {\it Recent Progress in Relativistic Astrophysics}, 6-8 May 2019 (Shanghai, China).}

\abstract{The search for periodic signals from blazars has become an actively pursued field of research in recent years. This is because periodic signals bring us information about the processes occurring near the innermost regions of blazars, which are mostly inaccessible to our direct view.  Such signals provide insights into some of the extreme conditions that take place in the vicinity of supermassive black holes that lead to the launch of the relativistic jets. In addition,  studies of characteristic timescales in blazar light curves shed light on some of the challenging issues in blazar physics that includes disk-jet connection, strong gravity near fast rotating supermassive black holes and release of gravitational waves from binary supermassive black hole systems. However, a number of issues  associated with the search for quasi-periodic oscillations (QPOs) in blazars e.g., red-noise dominance, modest significance of the detection, periodic modulation lasting for only a couple of cycles and their transient nature, make it difficult to estimate the true significance of the detection. Consequently, it also becomes difficult to make meaningful inferences about the nature of the on-going processes. In this proceedings, results of study focused on searching for QPOs in a number of blazar multi-frequency light curves are summarized.  The time series analyses of long term observations of the blazars revealed the presence of year-timescale QPOs in the sources including OJ 287 (optical), Mrk 501 (gamma-ray), J1043+2408 (radio) and PKS 0219 -164 (radio). As likely explanations, we discuss a number of scenarios including binary supermassive black hole systems, Lense–Thirring precession, and jet precession.}


\begin{document}


\section{Introduction}
 Active galactic nuclei (AGN) are the most luminous sources in the Universe. As accreting  matter swirls inward to the supermassive black hole at its center\citep[][]{2019ApJ...875L...6E}, the gravitational potential energy of the material is processed into radiation energy resulting in a large output of high energy emission. As a result, the central core, mainly the accretion disk with black hole at the center, outshines the whole galaxy. A small fraction of AGN happen to eject powerful  jets aligned along the line of sight. These objects are known as blazars which are dominant sources of relativistically boosted non-thermal sources in the universe.   Blazars  consist of  flat-spectrum radio quasars (FSRQ), some of the most powerful sources, and  BL Lacertae (BL Lac) objects, cosmic sources of highest energy emission. As blazars are bright and \emph{visible} across cosmological distances, they could be very useful in mapping the large scale structure of the Universe. More recently, blazar TXS 0506+056 was linked to the first non-stellar neutrino emission detected by the IceCube experiment \citep{2018Sci...361..147I}.

  Blazars display flux variations on diverse spatial and temporal frequencies \citep[see][and references therein]{Bhatta2018b}.  The flux variability shown by AGN  mostly appears aperiodic, but  presence of quasi-periodic oscillations (QPO) in the multi-frequency blazar (both flat spectrum radio quasars and BL Lacertae objects) light curves has been frequently reported, especially owing to the availability of long term observations from various instruments in a wide range of frequency bands e.g optical radio and $\gamma$-ray \citep[see][and references therein]{Bhatta2016b,Gupta2018}. 
 As a result, the hunt for periodic signals in blazar light curves has become an active field of research: blazar OJ 287 is famous for its $\sim12$-year periodicity detected in its historical optical light curve \cite{Sillanpaa1988,Valtonen2011} and another famous source PG 1553+113 is found to show 2.18-year periodic modulations in its $\gamma$-ray observations from the Fermi/LAT, which was first reported in \cite{Ackermann2015}. In addition, several authors have reported existence of QPOs of various characteristic timescales in numerous blazar multi-frequency light curves including the 4 cases of  QPO detection by our research group \citep[see][and the references therein]{Bhatta2016b,Bhatta2017,Bhatta2018d, Bhatta2019}. 

QPO signals mostly likely originate at the innermost regions of blazars and propagate along the jets to reach us.  Therefore, the signals carry crucial information the processes occurring near the central engine.  Performing an in-depth time series analysis of blazar light curves can provide insights into the processes near the AGN central engines and thereby add to our current understanding of the physics of supermassive black holes and origin relativistic jets. In this proceedings, I summarize our results of study of QPO analysis in blazars, and also discuss some of  the methods that are mostly used to detect possible periodic components in blazar multi-frequency light curves and  establish significance of the detection using Monte Carlo (MC) simulations.  Besides, some of the AGN scenarios that could possibly result in QPOs are explored.
 
\section{ Analysis Methods}
Some of the popular methods of time series analysis that are employed to search for periodicity in astronomical observations are discussed below.
\subsection{Discrete Fourier Periodogram} 
For a given time series $x(t_{j})$ sampled at times $t_{j}$ with $j = 1, 2,.., N$, the discrete Fourier power at a temporal frequency $\nu$ is estimated using the relation
\begin{equation}
P\!\left(\nu \right)=\frac{2\,T}{\left(N \bar{x} \right)^{2}} \, \left | \sum_{j=1}^{N}x\!\left( t_{i} \right ) \, e^{-i2 \pi \nu t_{j}} \right |^{2} \, ,
\end{equation}
where $T$ and $\bar{x}$ represent the total duration of the series, and the mean flux of a variable source, respectively \cite{Uttley2002}.  If the observations sampled at discrete time step $\Delta \rm t$ span a total duration of $\rm T$, the powers are computed at the evenly spaced log-frequency grid between the lowest frequency  $1/\rm T$ to the highest $1/(2\Delta t)$, also known as  Nyquist frequency. Thus computed discrete power is often known as discrete Fourier periodogram (DFP). The distribution of DFP  over the considered temporal frequencies reveal variability power of the source light curve at the corresponding timescales, and thereby provide crucial information about the underlying variability structures and dominant timescales. In particular, if a significant periodogram peak centered around a frequency is observed, it is  most likely  due to the oscillations in the data that repeat after the characteristic timescale corresponding to the peak frequency. The computational simplicity makes the method first choice among astronomers searching for periodic signals in astronomical observations. However, the results of the method are largely affected by the irregularities and gaps in the data.

\subsection{Lomb-Scargle Periodogram} 
While the DFP of a unevenly sampled finite light curve can produce many artifacts which could be mistaken as  periodic components, the Lomb-Scargle Periodogram \cite[LSP;][]{Lomb1976,Scargle1982} is considered to be an improved method to compute the periodograms.  For an angular frequency, $\omega$, the LSP is given as
\begin{equation}
P=\frac{1}{2} \left\{ \frac{\left[ \sum_{i}x_{i} \cos\omega \left( t_{i}-\tau \right) \right]^{2}}{\sum_{i} \cos^{2}\omega \left (t_{i}-\tau \right) } + \frac{\left[ \sum_{i}x_{i} \sin\omega \left( t_{i}-\tau \right) \right]^{2}}{\sum_{i} \sin^{2}\omega \left( t_{i}-\tau \right)} \right\} \, ,
\label{modified}
\end{equation}
where $\tau$ is given by
\begin{equation}
\tan\left( 2\omega \tau \right )=\frac{\sum_{i} \sin\omega t_{i}}{\sum_{i} \cos\omega t_{i}} \, .
\end{equation}

 The method attempts a fit to sine waves of the form $X_{f}(t)= A \cos\omega t +B \sin\omega t$  to the data. Therefore it  is less affected to the gaps and irregularity in the sampling of data when compared to DFP.  Besides, the least-square fitting process enhances the sinusoidal component in the periodogram. This makes the method more efficient in detecting periodic signals in light curves.  Usually the periodogram is computed for the minimum and maximum frequencies of, $\nu_{min} = 1/T$, and  $ \nu_{max}=1/(2 \Delta t)$, respectively. The total number of frequencies considered $N_{\nu}$ also plays an important role in the evaluation of the periodogram. It is empirically given as $N_{\nu}=n_{0} T \nu_{max} $,  where $n_{0}$ can be chosen in the range of $5-10$  \citep[see][]{VanderPlas2018}.  As with the DFP, a distinct peak in the LSP signifies possible presence of a underlying periodic signal.

\subsection{Weighted Wavelet z-transform}
 The DFP and LSP are efficient in detecting periodic signals especially when the signals span the entire observation duration. However, in real astronomical systems, periodic oscillations may develop and evolve in frequency and amplitude over time, hence quasi-periodic signals. In such cases, time series analysis based on the wavelet transforms are much more useful. Wavelet methods attempt to fit sinusoidal waves that are localized in both time and frequency domains.  In the context of astronomical observations where data are full of irregularities and gaps, weighted wavelet z-transform (WWZ) is one of the methods widely applied.  In this method, the wavelet functions can be viewed as weighted projections on the trial functions given as
\begin{equation}
 \phi_{1}\!\left( t \right) =\mathbf{1}\!\left( t \right)  \text{,} \ \ \ \phi _{2}=\cos\left[ \omega (t-\tau ) \right]  \ \ \ \text{and} \ \ \ \phi _{2}=\sin\left[ \omega (t-\tau ) \right],
\end{equation}
which further are weighted with a weight function $w_{i}=e^{-c \, \omega^{2}\left( t_{i}-\tau \right)^{2}}$, where $c \sim 0.0125$ is a fine tuning parameter.  Then WWZ power can be written in terms of weighted variations of the data,  $V_{x}$ and $V_{y}$, given as
\begin{equation}
\label{wwz}
WWZ=\frac{\left ( N_{eff}-3 \right )V_{y}}{2\left ( V_{x}-V_{y} \right )},
\end{equation}
and where $N_{eff}$ is the effective number of the data points \citep[see][for further details]{Foster1996}. When WWZ analysis is performed, the distribution of color-scaled WWZ power of the source in the time-period\footnote{Note that the time and the timescale corresponding to a temporal frequency are two different measures.} plane can reveal QPOs \citep[e.g.][]{Bhatta2017,Bhatta2016b}.  In case where the data contains a variable period  the maximum powers are seen to ``meander'' on the time-period plane. 

\subsection{Epoch folding} 
In contrast to the DFP, LSP and WWZ, which represent frequency domain based methods of time series analysis, epoch folding \cite[see][]{Leahy1983,Davies1990,Davies1991} is a method which is based on time domain.  The method has several advantages over traditional methods: it is capable of detecting the periodic signals of any shapes, it is relatively less affected to the uneven sampling and gaps in the observations, and it is free of all the artifacts associated with frequency domain analysis such as aliasing and red-noise leak. For periodicity tests, a given light curve is folded on a range of trial periods and phase bins,  and consequently $\chi ^{2}$ is computed according to
\begin{equation}
\chi ^{2}=\sum_{i=1}^{M}\frac{\left (x_{i}-\bar{x}\right )^2}{\sigma_{i}^{2}}, 
\label{chisq}
\end{equation}
\noindent where $x_{i}$ and $\sigma_{i}$ represent the mean and standard deviation, respectively, of each of M phase bins. In the absence of any periodic signal in the data, or the observations distributed are as Gaussian noise, we find $\chi ^{2}$ $\sim M$. However in the case when a light curve contains a periodic signal, $\chi ^{2}$ at the characteristic timescale becomes large and significantly different from it's mean value \cite[for details refer to][]{Larsson1996}. 
 $\chi ^{2}$ values are computed  for the trial periods in the range of suspected timescales. The  pulse profiles obtained for the trial periods are tested using  Equation \ref{chisq}.   Usually, the maximum $\chi^2$  deviation  at a timescale signifies the underlying most probable period in the data. 

\subsection{Discrete auto-correlation function}
The discrete auto-correlation function (ACF)  is another time domain based method frequently applied to search periods in astronomical time series. The method is based on the discrete correlation function (DCF), a method less affected to the sampling irregularity in the observations.  The unbinned DCF can be computed as,
\begin{equation}
UDCF_{ij}=\frac{\left ( x_{i}-\bar{x} \right )\left ( y_{j}-\bar{y} \right )}{\sqrt{\left ( \sigma _{x}^{2} -e_{x}^{2}\right )\left ( \sigma _{y}^{2} -e_{y}^{2}\right )}}
\label{UDCF}
\end{equation}
where $\bar{x}$ and $\bar{y}$ represent mean values of the two time series and  $\sigma^{2}$, $e^{2}$ correspond to variance and uncertainties in the light curves, respectively \cite[see][]{EK88}.  When these discrete pairs are binned in a suitable time lag bin,  the mean DCF of the  bin including the M pairs  is expressed as, 
\begin{equation}
DCF(\tau )=\frac{1}{M}UDCF_{ij}
\label{DCF}
\end{equation}
However, in this method the sampling distribution of DCF are often highly skewed such that the measure of DCF uncertainties represented by sample variances can be unreliable. In such a case, the  DCFs can be z-transformed (i.e., ZDCFs) so that the distribution closely resembles normal distribution, and then standard deviation provides a robust  estimation of ZDCF uncertainties \cite[For details see][]{Alexander2013}.  Now to compute ACF we calculate ZDCF using the same light curve in place of two time series data. If a time series observations contain periodic signals, the distribution of ACF over the time lag also shows periodic oscillations  \cite[see][and the reference therein]{Bhatta2018a}.

In all the of methods described above, the half width at the half maximum of the Gaussian fit to the central peak, that signified the presence of QPO, was used as a measure of the uncertainty in the detected period.

\subsection{Significance Estimation and Monte Carlo Simulation}

An observed light curve $x(t_i)$, sampled at discrete $t_i\ (\text{ where}\ i=1 \text{ through}\ N$) times can be viewed as the a continuous light curve $x(t)$ multiplied with a window function $w(t)$ such that

\begin{equation}
w(t) =
    \begin{cases}
            1, &         \text{when sampled } \\
            0, &         \text{otherwise}.
    \end{cases}
\end{equation}
 The Fourier transform of such a function can be expressed as the  Fourier transform of continuous function convolved with that of the window function, i.e.,
\begin{equation}
F(\nu)=X(\nu)^*W(\nu).
\label{Fourier}
\end{equation} 
From the above relation, it is clear that the true underlying PSD of the observed light curve is in fact distorted by the effects of sampling pattern, i.e., window function.  Besides,  blazar light curves which are mostly dominated by power-law type noise can sometimes show  a few cycles of periodic flux modulation, especially in the low-frequency domain \citep[see][for the discussion]{Press1978,Bhatta2017,Bhatta2018c,2019MNRAS.482.1270C}.  

Power response method  \citep[PSRESP;][]{Uttley2002}, one of the methods extensively applied in the characterization of  PSDs of AGN, incorporates  these effects in its significance estimation method through the use of a large number of simulated light curves that mimic several properties of real observations, such as mean, standard deviation, sampling pattern and duration.  Simulations of  light curves can be performed following the Monte Carlo (MC) method described in \cite[][]{Timmer1995}. 
First, the source periodogram is modeled with a power-law PSD of the form $P(\nu) \propto \nu ^{-\beta}+C$; where ${\nu}$ and ${\beta}$  represent temporal frequency and spectral index, respectively.  For a light curve spanning the total duration of time T  and with a mean flux of $\bar{x}$, Poisson noise $C$ level can be given by
\begin{equation} 
 C=\frac{2T}{N^2\bar{x}^2}\bar{\Delta{x^2}},
\end{equation} 
where ${\Delta{x}}$ represent the flux uncertainties.   In order to obtain a measure of significance of the periodic feature,   $\chi ^2$ values are computed for the observed and the simulated light curves as following
\begin{equation}
\chi_{\rm obs}^{2}=\sum_{\nu _{\rm min}}^{\nu _{\rm max}}\frac{\left [P_{\rm obs}\left ( \nu  \right )- \overline{P_{\rm sim}\left ( \nu  \right )} \right ]^{2}}{\Delta \overline{P_{\rm sim}\left ( \nu  \right )}^{2}} \quad\text{and} \quad  \chi_{\rm dist, i}^{2}=\sum_{\nu _{\rm min}}^{\nu _{\rm max}}\frac{\left [P_{\rm sim, i}\left ( \nu  \right )- \overline{P_{\rm sim}\left ( \nu  \right )} \right ]^{2}}{\Delta \overline{P_{\rm sim}\left ( \nu  \right )}^{2}},
\label{chi}
\end{equation}
\noindent where $P_{\rm obs}$ represents observed periodogram, and  $P_{\rm sim, i}$, $ \overline{P_{\rm sim}\left ( \nu  \right )}$ and $\Delta \overline{P_{\rm sim}\left ( \nu  \right )}$ stand for periodogram of each simulated light curve, the mean periodogram and standard deviation of all the  simulated periodograms, respectively.  Then ratio of the number of $\chi^{2}_{i}$s greater than $\chi^{2}_{obs}$ to the total number of $\chi^{2}_{i}$s in all simulations provides a measure of the probability than can be used to quantify the goodness of the fit for a given model. These ratios are computed for various trial shapes and slope indexes such that the best fitting PSD model has highest probability \citep[see][for further details]{Bhatta2016,Chatterjee2008}. Here it is important to make a distinction between local and global significance. The local significance tells us how likely a periodic signal at a particular frequency is significant; whereas the global significance accounts for the fact that we do not have an {\it a priori} knowledge of the location of the frequency at which the signal might occur,  and hence make considerations for all the periodogram frequencies \cite[see][]{Bhatta2017}. 

\section{Results}

Under a broader program of characterizing long term, multi frequency statistical properties of blazar sources, we performed time series analysis on a sample of blazars. Below we present particularly the periodicity analysis on four well-studies sources. In addition, the summary of the results of the analyses are presented in Table \ref{table:1} that lists source names, red-shift of the sources, the waveband of the observations, total observation duration in approximately in years, the number of observation points, the analysis methods applied to derive the periods, the source rest-frame periods approximately in days, and the number of cycles, approximated to nearest whole number, observed in the light curve, and the references for the previously published works in column 1, 2, 3, 4, 5, 6, 7,  8 and 9, respectively. 

\begin{itemize}

\item Mrk 501: Decade-long  Fermi/LAT observations (100 MeV-300 GeV) of the famous TeV blazar Mrk 501 were analyzed using four widely known methods:  Lomb-Scargle periodogram, weighted wavelet z-transform, epoch folding, and z-transformed discrete auto-correlation function. The analyses revealed a underlying sub-year timescale (332 d) $\gamma$-ray QPO that persisted nearly 7 cycles.  Applying PSRESP analysis  the significance of the detection was found to be above 99\% against possible spurious detection \citep[for details see][]{Bhatta2019}. \\

\item J1043+2408:  A $\sim$ 560-d QPO was detected  in the radio (15 GHz) observations of the BL Lac source J1043+2408.  Multiple methods of time series analysis, such as  epoch folding, Lomb-Scargle periodogram, and discrete auto-correlation function, were carried out on the source light curve. All three methods consistently found the repeating signals which were identified as being identical in periods within the uncertainties.  In order to estimate the significance of the detection, a large number of Monte Carlo simulations were performed. This resulted in a significance of 99.9\%  99.4\%   local and global, respectively. \citep[for details see][]{Bhatta2018d}. \\

\item PKS 0219-164: The long-term variability properties of the blazar PKS 0219-164 were investigated utilizing the 15 GHz radio observations from the period 2008–2017.  The methods Lomb-Scargle periodogram and weighted wavelet z-transform  detected a strong signal of 270-day QPO along with possible harmonics of 550 and 1150  days periods. The  statistical significance, both local and global, was estimated above   99\%  over underlying red-noise processes \citep[for details see][]{Bhatta2017}. \\

\item  OJ 287: we performed time series analysis of $\sim$ 9.2 year-long, densely-sampled optical light curve of the BL Lac OJ 287.  Both the LSP and the WWZ methods provided evidence for the possible $\sim$  400 d QPOs in the source light curve with a high significance of $>$ 99\%.  In addition, hint for a possible harmonic of $\sim$  800 d period was also found \citep[for details see][]{Bhatta2016b}.

\end{itemize}

\begin{table}[H]
\caption{The list of the blazars in different wavebands that were detected to show periodic oscillations.}
\label{table:1}
\centering
\begin{tabular}{lclclcccl}
\toprule
Source	& Red-shift	& Waveband&Obs. dur. (Yrs.)& Npts& Methods& Period (d)& No. of cyc.& Ref.\\
\midrule
OJ 287		&0.306& optical	& 9.2&1338		& 1, 2& 410$\pm38$&8&\cite{Bhatta2016b}\\
PKS 0219 -164& 0.7&radio & 8.4&345	& 1, 2& 160$\pm26$&11&\cite{Bhatta2017}\\
J1043+2408&  0.563446&radio & 10.5&524	& 1,  3, 4& 360$\pm49$ 	&6	&\cite{Bhatta2018d}\\
Mrk 501	&0.034	& $\gamma$-ray	& 10.0&461		& 1, 2, 3, 4& 320$\pm17$&10&\cite{Bhatta2019}\\
\bottomrule
\end{tabular}\\

 \begin{tablenotes}
           \item Note: 1$\rightarrow$Lomb-Scargle periodogram, 2$\rightarrow $weighted Wavelet transform, ACF$\rightarrow$ auto-correlation function, and EF$\rightarrow$epoch folding
    \end{tablenotes}
\end{table}



\section{Discussion and Conclusions}

We reported the presence of year timescale QPOs in 4  blazars in different wave bands by applying multiple methods of time-series analysis. It should be pointed out that the fact that all 4 sources in our studies are of BL Lac type blazars should not give impression that QPOs are less frequent in FSRQs. In fact,  the study of the possible differences between the types of QPOs detected in these two sub-classes of the blazars is yet to be carried out.

Time domain analysis of multifrequency blazar light curves provides an excellent tool to investigate the physical processes occurring at the innermost  regions of blazar, which otherwise are hardly accessible to current instruments. Such an approach can offer important insights into a number of challenging issues including nature of space-time around fast spinning supermassive black holes (SMBH),  disk-jet connection and release of gravitational waves from the binary supermassive black hole systems. A number of possible scenarios can be linked to  the observed year timescale for QPOs in blazars,  some of which are discussed below.

First, from the observed period ($P_{obs}$), the corresponding period in the source rest frame ($P$) at the redshift of $z$  is estimated  using the relation
\begin{equation}
P=P_{obs}/(1+z).
\end{equation}
 As a simplest case, the observed period might represent Keplerian period $\tau_k$ around central black hole  - such as a bright hotspot revolving around the black hole. In such case, the timescale can be used to estimate the  radius of the orbit using the relation
\begin{equation}
\tau_k=0.36 \left ( \frac{M}{10^9M_\odot} \right )^{-1/2}\left ( \frac{a}{r_g} \right )^{3/2} \rm days,
\end{equation}
where $a$ is the length of the semi-major axis of the elliptic orbits. Assuming circular orbits, for a typical black hole of mass of $10^9 M_\odot$ the radius of the Keplerian orbit for a year timescale can be estimated  to be a few tens of gravitational radius ($r_g$).  Similarly, the timescale might also represent the period of the supermassive binary black hole system (SMBBH) \citep[e.g.,][]{Valtonen2011}. In such a system, elliptical orbits are gradually  turned into circular orbits by dynamical friction that acts over the long course of merging. Therefore, the radius of the Keplerian orbit can be calculated using the relation,
  \begin{equation}
r=9.5\times 10^{-5}\left ( \frac{M}{10^9M_\odot } \right )^{1/3}\tau _k^{2/3} \rm \ pc
 \end{equation} 
For a typical SMBBH system, the total mass of the system can fairly be assumed to be $1\times10^9 M_\odot $. Inferred from the year timescale periodicity, the black holes can be revolving around each other at a separation of a few  milli-parsecs.  Such close binary system undergo  orbital decay due to emission of gravitational waves.   For the binary mass ratios in the rage 0.1--0.01 \cite[see e.g.,][]{Begelman1980,Sillanpaa1988}, the orbital decay timescale in the GW-driven regime as estimated  using the relation
 \begin{equation}
\tau_{insp}=3.05\times 10^{-6} \left ( \frac{M}{10^9M_\odot} \right )^{-3}\left ( \frac{a}{r_g} \right )^4 \rm years,
\end{equation}
can be less than a thousand years \cite[see][]{Peters1964}.  In such case we should expect that the gravitational coalescence of the system within a few centuries along with the emission of low frequency ($\sim10^{-2}$ $ \mu$Hz) gravitational waves.   But on the other hand, if the orbits in (SMBBH) are significantly elliptical, the secondary BH periodically might perturb the primary jet at periastron causing magnetohydrodynamic instabilities. These events inducing particle acceleration via magnetic reconnection events can release electromagnetic emission via synchrotron and inverse Compton emission \citep[see][]{Tavani2018}. Moreover, the secondary black hole may perturb the disk of primary black hole such that the jets of the primary starts precessing across the line of sight to result in the apparent QPO \citep[see][]{1997ApJ...478..527K}.

 In a different scenario, the observed year timescale QPOs  could be associated with the instabilities taking place at the accretion disks. For example,  oscillations developed in globally perturbed thick accretion can also give rise to QPOs  and their fundamental frequency of the oscillation can be approximated as,  
   \begin{equation}
  f_{0}\approx 100\left ( \frac{r}{r_g} \right )^{-3/2}\left ( \frac{M}{10^8M\odot } \right )^{-1} \ \rm day^{-1}
 \end{equation}
 \citep[see][]{Liu2006,An2013}.  
Similarly, QPOs can be linked to the precession of warped accretion disks around Kerr black holes residing at the hearts of blazars,  known as the Lense-Thirring precession  \citep[e.g.][]{Stella1998,Motta2011}. The timescale of such precession,  $\tau_{LT}$, can be expressed in terms of the distance from the BH $r$ and  the mass of the central black hole $M$ as given by
 \begin{equation}
\tau_{LT}=0.18 \left ( \frac{1}{a_s} \right )\left ( \frac{M}{10^9M_\odot} \right )\left ( \frac{r}{r_g} \right )^{3} \rm days,
\end{equation}
\noindent where $a_s$ represent dimensionless spin parameter. For a maximally spinning ($a_s=0.9$) central black hole  with a mass ${10^9M_\odot}$, a timescale of 360 days places the emission region around $\sim 12\ r_g$.  Furthermore, in the case  where the accretion disk is strongly coupled to the jet \cite[e.g., in][]{Blandford1982} (see also \cite[][]{Bhatta2018c}), the precession of the disk can drive the jets precession. In fact, recent studies of 3D general relativistic MHD simulations ( \citep[see][]{Liska2018}) based on the similar scenario has shown that the jets exhibit precession with a period of $\sim 1000\ r_g/c$, which for a $10^9M_\odot$ black hole turns out to be of the order of a few years.  Moreover, the modulations arising near the central engine could propagate down the jet where the emission and the timescales are altered  due to relativistic effects such that true periodic timescales at the BH could be longer than the observed according to the relation
\begin{equation}
P=\delta /(1+z)P_{obs} \ \ \text{and} \ \  \delta =(\Gamma \left ( 1-\beta cos\theta \right ))^{-1}
\label{Doppler_fact}
\end{equation}
where  for $\Gamma$, $\beta=v/c$ and $\theta$ are the bulk Lorentz factor, relativistic speed and  angle  to the line of sight.
 Equation \ref{Doppler_fact}  also suggests that QPOs in blazar could be related to  the periodic swings in the angle between the emission regions and the line of sight. For example, emission regions moving along helical path of the magnetized jets can lead to the periodic changes in the viewing angle \cite[e.g.,][]{Camenzind92}.   This way the  Doppler boosted emission from the jet  can be modulated periodically resulting in QPOs. In such cases, the rest frame flux (${F}'_{{\nu'}}$) is related to the observed flux ($F_{\nu}$) through the equations
  \begin{equation}
F_{\nu}(\nu)=\delta(t)^{3+\alpha}{F}'_{{\nu}'} (\nu)   \quad\text{and}\quad      \delta(t)=1/\Gamma \left ( 1-\beta cos\theta(t)  \right ).
\label{flux}
\end{equation}
Following the above argument, intrinsic flux remaining unchanged,  the ratio of the observed  flux modulation for a given change in the angle $\Delta \theta$ can be obtained from the relation
 \begin{equation}
 \Delta logF=-\left ( 3+\alpha  \right )\delta \Gamma \beta sin\theta \Delta \theta ,
 \end{equation} 
 As an illustration, for blazars having typical radio spectral index ($\alpha=0.6$) and viewing angles in the $1-5 ^o$ range, a slight change in the viewing angle e.g., $\sim 1.5^o$, is sufficient to produce observed maxima twice as bright as minima in the light curves \citep[refer to Figure 4 in][]{Bhatta2018d}.

Finally it is concluded that, In blazars we observe QPOs  from mainly the jet emission that might be driven by perturbations near the black hole.   In such context, recurring gravitational perturbation in SMBBH systems driving the QPO production mechanism in jets serve as a most natural model.   However, several issues linked with searching for QPOs in blazars such as red-noise dominance, modest significance of the detection and transient nature of the QPOs hinder the efforts to definitively characterize them. Therefore, the future studies should be directed to more comprehensive multi-frequency time series analysis and collective discussion on the topic.
\vspace{6pt} 



\funding{This research was funded by Nardowe Centrum Nauki (NCN) grant UMO-2017/26/D/ST9/01178}

\conflictsofinterest{The authors declare no conflict of interest} 

\abbreviations{The following abbreviations are used in this manuscript:\\

\noindent 
\begin{tabular}{@{}ll}
AGN & Active Galactic Nuclei\\
ACF & Auto-correlation function\\
BL Lac & BL Lacertae object\\
BH & Black hole\\
FSRQ & Flat Spectrum Radio Quasar\\
LSP & Lomb-Scargle Periodogram\\
Monte Carlo  & MC\\
PSD&Power spectral density \\
QPO&Quasi-periodic oscillation \\
SMBH& Supermassive black hole\\
SMBBH& supermassive binary black hole\\
WWZ& Weighted Wavelet z-transform\\
\end{tabular}}


\reftitle{References}



\sampleavailability{Samples of the compounds ...... are available from the authors.}


\end{document}